\documentstyle[aps,pre,preprint,amsfonts]{revtex}
\tightenlines
\begin{document}
\draft
\title{Upper Bounds for the Critical Car Densities in Traffic Flow Problems}
\author{H. F. Chau}
\address{School of Natural Sciences, Institute for Advanced Study,\\
Olden Lane, Princeton, NJ 08540, USA}
\author{P. M. Hui and Y. F. Woo}
\address{Department of Physics, The Chinese University of Hong Kong,\\
Shatin, New Territories, Hong Kong}
\date{June 30, 1995}
\preprint{IASSNS-HEP-95/07; adap-org:9502002}
\maketitle
\mediumtext
\begin{abstract}
In most models of traffic flow, the car density $p$ is the only free
parameter in determining the average car velocity $\langle v \rangle$.
The critical car density $p_c$, which is defined to be the car
density separating the jamming phase (with $\langle v \rangle = 0$) and
the moving phase (with $\langle v \rangle > 0$), is an important physical
quantity to investigate. By means of simple statistical argument, we show
that $p_c < 1$ for the Biham-Middleton-Levine model of
traffic flow in two or higher spatial dimensions. In particular, we show
that $p_{c} \leq 11/12$ in 2 dimension and $p_{c} \leq 1 - \left(
\frac{D-1}{2D} \right)^D$ in $D$ ($D > 2$) dimensions.
\par\bigskip\noindent
Keywords: Biham-Middleton-Levine Model, Car Jamming Phase, Cellular Automata,
Critical Car Density, Traffic Flow
\end{abstract}
\medskip
\pacs{PACS numbers: 89.40.+k, 05.60.+w, 64.60.Ak, 89.50.+r}
Recently, there has been much interest in the study of traffic flow
problems within the context of cellular automaton (CA) models.	This
approach has the advantages of being simple and computationally less
intensive.  Nagel and Schreckenberg \cite{Nagel} introduced a stochastic
discrete
automaton model in one dimension (1D) to study the transition from laminar
traffic flow to start-to-stop-waves as the car density increases.
Variations of the 1D model have been extensively studied
\cite{Naga1,Fukui,Naga2,Chung1}.
Biham, Middleton, and Levine \cite{BML} (henceforth referred to
as BML) introduced a
simple two dimensional (2D) CA model with traffic lights and studied the
average velocity of cars as a function of their concentration.	In the BML
model, each site on a 2D square lattice may take on an east-bound car (i.e.,
a car traveling in positive x-direction) and a north-bound car (i.e., a
car traveling in positive y-direction) both with probability $p/2$.
Otherwise, the site is empty with probability $1-p$.
The probability $p$ thus corresponds to the car
density.  Each east-bound car moves at even time steps by one site provided
that the neighboring site to the east is empty.  Similarly, each
north-bound car moves at odd time steps by one site provided that the
neighboring site to the north is empty.  BML found that, using numerical
simulations, the average car velocity in the long time limit vanishes when
the car density is higher than a critical value $p_{c}$.  The BML model
has been generalized \cite{Naga3,Naga4,Cuesta,Chung2}, for example, to include
effects
of two-level crossing (overpasses), faulty traffic lights, accidents, and
anisotropy in car densities in east and north bound directions.
Generalization of the model to other spatial dimensions is straight forward.
``Traffic" models in higher dimensions, though may not be relevant to cars
on the road, may be related to problems in information network and
telecommunication.

Nagatani \cite{Naga3,Naga4} has proposed a mean field theory for the average
car velocity. The theory has recently been criticized by Ishibashi and
Fukui \cite{Ishibashi}. They suggested that no complete traffic jam occurs
unless $p = 1$, in contrast to simulation results and other
mean field approximations. They also speculated that the values of $p_c$
obtained in numerical experiments may be results of finite size effects and
the use of periodic boundary conditions \cite{Ishibashi}. It is therefore
desirable to derive bounds
for $p_{c}$ in various spatial dimensions.  In particular, if the upper
bound of $p_{c}$ is less than unity, then there exists a {\em range} of car
densities corresponding to a complete traffic jam.  In what follows, we
argue that there exists an upper bound for $p_{c}$ less than unity whenever
the spatial dimension of the lattice is greater than one.  In 2D,
we have $p_{c} \leq 11/12$ and we also give upper bounds for $p_{c}$ for
spatial dimensions $D > 2$.

Consider the BML traffic flow problem on a 2D square lattice.  Instead of
looking at how the traffic flows, we concentrate on how the traffic jams.
We note that the 2D square lattice can be partitioned, for example, by a
collection of inverted-L-shaped elements as shown in Fig. 1.  Each element
consists of 3 sites.  For example, the element at the site $(i,j)$ contains
a collection of three sites $L_{i,j} \equiv \{(i,j),(i-1,j),(i,j-1)\}$.
Thus, we can partition the 2D lattice by $\{ L_{3i+j,j} : i,j \in {\Bbb Z}
\}$ (see Fig.1).   Obviously, this is not the only way to partition
the lattice. As we shall see, different partitions may lead to different
upper bounds.	It is, however, suffice for our purpose to consider this
simple partition.  The important point is that the elements can fill the
underlying lattice with a density of non-zero measure.

Consider the case of a full lattice of cars, i.e., $p=1$, which trivially
corresponds to a jamming phase.  Consider, say, the element at $(i,j)$.
The car at site $(i,j)$ can be removed without changing the system to a
moving phase if the cars at sites $(i-1,j)$ and $(i,j-1)$ are north-bound
and east-bound respectively.  Assume the cars to have equal probability of
being east or north bound, the probability of having the car at the middle
of the inverted-L-shaped element removed without changing the system to a
moving phase is $(1/2)^{2}$. Since each of the inverted-L-shaped element
contains three sites, so in the limit of infinite system size, we can
almost surely remove $(1/3) \times (1/2)^{2} = 1/12$ of the cars in the
system for {\em any} initial configuration of a full lattice of cars
without changing the system from a jamming phase to a moving phase.  This
suggests an upper bound of $p_c \leq 11/12$ in 2D. Conversely, let us consider
the case for an arbitrary initial configuration with car concentration $p =
11/12$. A car moves into an empty site $(i,j)$ in the next time step or two
whenever there is a north-bound car located at $(i,j-1)$ or an east-bound car
located at $(i-1,j)$ at the present moment. The probability for this to occur
is given by $q = \frac{11}{12}\ \frac{1}{2} + \frac{11}{12}\ \frac{1}{2} -
\left( \frac{11}{12}\ \frac{1}{2} \right)^2 =
\frac{11}{12}\times\frac{37}{48}$. Therefore, in the mean field approximation,
the expected number of sites at a given empty site $(i,j)$ can move is given by
\begin{eqnarray}
\langle n\rangle & \approx & (1-q)\times 0 + q(1-q)\times 1 + q^2 (1-q)\times 2
+ \cdots + q^k (1-q)\times k + \cdots \nonumber \\ & = & \frac{q}{1-q}
\mbox{~.}
\end{eqnarray}
So on the average, an empty site will be ``blocked'' after moving over
$\langle n\rangle \approx 407/169 \approx 2.4$ sites. This is less than the
typical separation between two empty sites (which equals $\sqrt{12}\approx
3.4$). Therefore, it is almost certain that a
system with car concentration $p = 11/12$ will evolve eventually to a jamming
state. Our result should be contrasted with that of \cite{Ishibashi}, in which
it was found that complete jam occurs only when $p = 1$.  We believe that
the discrepancy is due to the fact that
Ishibashi and Fukui have incorrectly assumed that all car configurations are
independent of each other. The upper bound on $p_c$ that we obtained here is
consistent with all the numerical experiments (Biham {\em et al.} found that
$p_c \approx 0.32$ for $512\times 512$ lattice \cite{BML}, while a
finite size scaling analysis by Nagatani showed that $p_c = 0.42$
\cite{Naga3}). We also note that our
argument above is true for {\em any} boundary conditions.  And
a more careful choice of partitions and elimination procedure may lead to
a tighter bound \cite{FutureWork}.

Our argument can be readily generalized to anisotropic
distribution of north and east bound cars as long as the traffic problem is
not decoupled into independent 1D lanes.  In 1D, it is known that complete
traffic jam only occurs when $p=1$ \cite{BML}.
Let $q_{e}$ ($q_{n}$) be the
probability of finding an east-bound (north-bound) car at a site given that
there is a car at the site. We have $q_{e} + q_{n} = 1$, and the ratio
$q_{e}/q_{n}$ reflects the anisotropy.	In this case, our argument leads to
a bound in 2D
\begin{equation}
p_{c} \leq 1 - \frac{q_{e} q_{n}}{3} .
\end{equation}
Even in the 1D limit, either $q_{e}$ or $q_{n}$ vanishes and $p_{c} \leq
1$, which is consistent with the known results $p_{c} = 1$.

Our argument can be carried over to any spatial dimension $D > 2$.
A naive extension of the above partition to 3D may be that of introducing
elements with four sites.  For example, the element $L_{ijk}$ at site
$(i,j,k)$ contains the collection of sites $L_{ijk} \equiv
\{(i,j,k),(i-1,j,k),(i,j-1,k),(i,j,k-1)\}$.  However, one cannot partition a
3D cubic lattice with elements $L_{ijk}$.
Since
the choice of partition is quite arbitrary, we choose, for simplicity, to
fill the underlying lattice with hypercubic elements in $D$ dimension, each
of which contains $2^{D}$ sites.  Consider a full lattice of cars.  The
probability of removing a car from a hypercube without changing the jamming
character is $(1 - \frac{1}{D})^{D}$.  Thus for $D > 2$, we have an upper
bound for $p_{c}$ given by
\begin{equation}
p_{c} \leq 1 - \frac{1}{2^{D}} \left( 1 - \frac{1}{D} \right)^{D} = 1 -
\left( \frac{D-1}{2D} \right)^{D} .
\end{equation}
In 3D, this gives $p_{c} \leq 26/27$.
For anisotropic distributions in different directions, we have
\begin{equation}
p_{c} \leq 1 - \frac{1}{2^{D}} \prod_{\ell = 1}^{D} (1 - q_{\ell}),
\end{equation}
where $q_{\ell}$ is the probability of finding a car at a site moving in
the $\ell$-th direction provided that there is a car at the site, and thus
$\sum_{\ell}^{D} q_{\ell} = 1$. As a result, in the infinite dimensional
limit, we find that $p_c \leq 1$ which is consistent with the mean field
theory result that $\langle v\rangle = 1 - p$ where $p$ is the probability
of finding a car in a site.

In summary, we argued that the critical car density has an upper bound less
then unity in traffic flow problems in $D > 2$.
\acknowledgements{
This work was initiated while one of us (HFC) was visiting the Chinese
University of Hong Kong under the support of a Direct Grant for Research
1994-95 at CUHK.  Work at the Institute for Advanced Study was supported by
DOE grant DE-FG02-90ER40542.
}

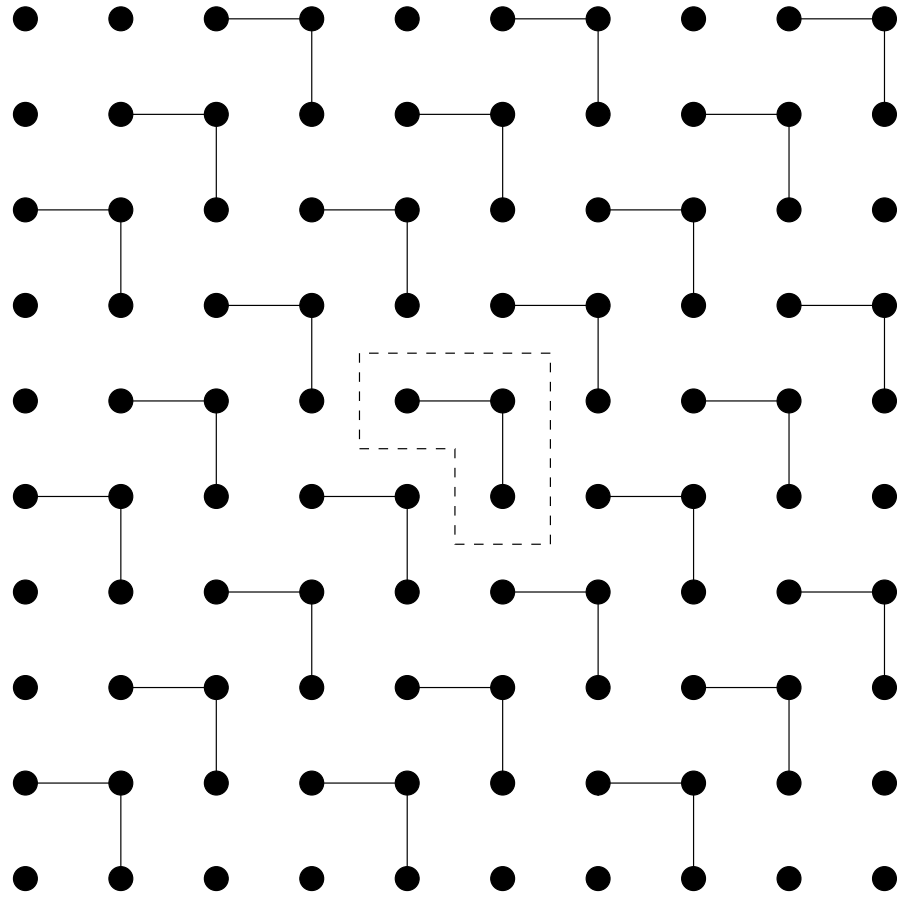
\begin{figure}
\caption{Partition of a two dimensional square lattice by elements
containing three sites.  The dashed line outline one of the elements.}
\end{figure}
\end{document}